\documentclass[twocolumn,showpacs,nofootinbib,amsmath,amssymb,prd]{revtex4}

\usepackage{hyperref}
\usepackage{amsfonts}
\usepackage{graphicx}

\begin{document}

\title{Conformal thin-sandwich puncture initial data for boosted black holes}

\author{Mark D. Hannam}
\email{marko@phys.utb.edu}
\affiliation{Center for Gravitational Wave Astronomy, University of Texas at Brownsville, 80 Fort Brown, Brownsville, TX 78520}
\author{Gregory B. Cook}
\email{cookgb@wfu.edu}
\affiliation{Department of Physics, Wake Forest University, Winston-Salem, North Carolina 27109}

\begin{abstract} 
We apply the puncture approach to conformal thin-sandwich black-hole initial data. We solve numerically the conformal thin-sandwich puncture (CTSP) equations for a single black hole with non-zero linear momentum. We show that conformally flat solutions for a boosted black hole have the same maximum gravitational radiation content as the corresponding Bowen-York solution in the conformal transverse-traceless decomposition. We find that the physical properties of these data are independent of the free slicing parameter.
\end{abstract}


\maketitle

\section{Introduction}

Numerical simulations of binary black-hole collisions require initial data. A black-hole collision scenario of current interest involves two black holes in orbit before they collide. As the black holes orbit, they emit gravitational radiation, and their orbits' radii slowly decrease. Eventually the two black holes are so close that no stable orbits exist, and they plunge together and merge. Gravitational radiation emission circularizes the orbits, so we expect the final orbits before merger to be almost circular. For many reasons long-term numerical evolutions are difficult, and therefore ideal initial data for a binary black-hole collision simulation would describe two black holes in an almost circular, slowly decaying orbit just prior to their merger, as well as all of the gravitational radiation that has built up during their slow inspiral. We do not yet know how to construct such initial data. However, a number of techniques exist to construct initial data for two boosted black holes, and to identify initial-data sets that exhibit some of the characteristics of circular orbits \cite{cook00,ggba,ggbb,tichy03,tichy04,yo04,cook04}. In this paper we outline a new procedure to construct initial data for multiple boosted black holes, and present solutions for single boosted black-hole spacetimes. 

A popular starting point for constructing binary black-hole initial data is the Bowen-York solution of the momentum constraint \cite{york79,bowen80}. This solution can be used to construct two black holes with arbitrary prescribed masses, linear momenta and spins. An effective-potential technique was developed by Cook \cite{cook94} to choose values of these parameters that correspond to quasi-circular orbits. This method also allowed a prediction of the location of the innermost stable circular orbit (ISCO) of the two black holes.

In Cook's original approach, the black holes were constructed on a two-sheeted topology, where the sheets were connected at the black-hole throats. The initial data was inversion symmetric between the two sheets. This approach allowed inner boundary conditions to be placed on the black-hole throats, and the remaining initial-value equation (the Hamiltonian constraint) was solved on only one sheet, with the region inside the throats (the second sheet) excised from the computational domain, avoiding all coordinate singularities. A later approach by Brandt and Br\"{u}gmann \cite{brandt97} removed analytically the coordinate singularity at the location (or ``puncture'') of each black hole, and made it possible to solve the Hamiltonian constraint without excising any region from the computational domain, and without any need for inner boundary conditions. Baumgarte \cite{baum00} used Cook's effective-potential method to construct initial data for binary black holes in quasi-circular orbits using the puncture approach. The predictions of the ISCO for inversion-symmetric and puncture Bowen-York data were in good agreement.

The Bowen-York solution arises from making certain choices of the freely specifiable quantities in the conformal transverse-traceless (CTT) decomposition of Einstein's equations \cite{york79}. These choices are made because they make the equations easier to solve, and not for any strong physical reasons. As such, we have cause to doubt that these data are astrophysically realistic and we seek alternatives to Bowen-York data that are more physically motivated.

A recent alternative decomposition of the initial-value equations is York's conformal thin-sandwich (CTS) decomposition \cite{york98}. The free quantities in the CTS decomposition are more closely linked to dynamics than those in the CTT decomposition. In particular, we can make choices consistent with a quasi-equilibrium spacetime, which is what we expect for two black holes in quasi-circular orbit \cite{cook00}.

Binary black-hole initial-data sets have been constructed using the CTS decomposition by a number of groups. Grandcl\'ement, Gourgoulhon and Bonazzola \cite{ggba,ggbb} solved a variant of the CTS equations using excision techniques, to construct conformally flat initial data. Cook and Pfeiffer \cite{pfeiffer02,cook04} also used excision techniques, but with improved inner boundary conditions. Yo, {\it et. al.} \cite{yo04} used a simplified version of the inner boundary conditions suggested in \cite{cook01}. 

It would be useful to be able to solve the CTS equations in the puncture approach \cite{baum03,hannam03}. Without the need for complex inner boundary conditions, the puncture approach is potentially easier to implement than excision. CTS-puncture binary black-hole initial data would provide a useful comparison with excision data results, and also allow evolution codes that employ punctures \cite{alcubierre03,brugmann04,imbiriba04} to take advantage of CTS data. 

The puncture approach was extended to the CTS equations in \cite{hannam03}. There it was pointed out that it will be difficult (and perhaps impossible) to construct CTS binary black-holes in quasiequilibrium in the puncture framework. It will be important to investigate the impact of this within quasi-circular orbits in CTS-puncture binary data, but we will not address that issue in this paper. For the moment we will confine ourselves to outlining a general procedure for constructing multiple boosted black-hole spacetimes, and apply our method to single boosted black holes. We will leave the question of orbits for a later publication.

Our starting point is a result by Laguna \cite{laguna04}, who found analytic solutions to an approximation to the conformal thin-sandwich puncture (CTSP) momentum constraint. These solutions describe a single conformally-flat black hole with non-zero linear or angular momentum, and reproduce the Bowen-York form of the conformal extrinsic curvature. Laguna's single-hole solutions provide an analytic result with which to test our numerical code, and also motivate the technique we employ for the full coupled CTSP system. This technique, which amounts to specifying a value for the shift vector at the punctures(s), allows us to construct initial-data sets for one or more black holes, each with non-zero linear momentum. 

In Section \ref{sec:cts} we review the conformal thin-sandwich decomposition, and in section \ref{sec:ctsp} outline its incorporation into the puncture framework. We consider Laguna's solution for a single black hole with non-zero linear momentum in Section \ref{sec:1BH}. We describe our numerical methods in Section \ref{sec:cvg}. We present numerical solutions of the full CTSP system for a single boosted black hole in Section \ref{sec:ctsp_BH}, and consider their physical properties.

\section{Conformal thin-sandwich decomposition}
\label{sec:cts}

In the ADM (3+1) decomposition of Einstein's equations \cite{adm,york79}, the ten Einstein equations are projected onto spacelike hypersurfaces, resulting in four constraint equations that the metric, $\gamma_{ij}$, and extrinsic curvature, $K_{ij}$, of each timeslice must satisfy, and six evolution equations that tell us how $\gamma_{ij}$ and $K_{ij}$ evolve to the next timeslice. In the 3+1 decomposition the spacetime metric can be written \begin{equation}
ds^2 = - N^2 dt^2 + \gamma_{ij} (dx^i + \beta^i dt) (dx^j + \beta^j dt). \label{eqn:3+1metric}
\end{equation} The lapse function $N$ gives the proper time between slices, $Ndt$, and the shift vector $\beta^i$ describes the coordinate drift between slices. Using these quantities one can calculate the extrinsic curvature using \begin{equation}
K_{ij} = - \frac{1}{2N} \left( \partial_t \gamma_{ij} - \bar{\nabla}_i \beta_j - \bar{\nabla}_j \beta_i \right).
\end{equation} The constraint equations on one slice are, in vacuum, \begin{eqnarray} 
\bar{R} + K^2 - K_{ij} K^{ij} & = & 0, \label{eqn:HC} \\
\bar{\nabla}_j K^{ij} - \gamma^{ij} \bar{\nabla}_j K & = & 0. \label{eqn:MC}
\end{eqnarray} The covariant derivative $\bar{\nabla}_i$ is taken with respect to the spatial metric, as is the Ricci scalar $\bar{R}$. The quantity $K$ is the trace of the extrinsic curvature. 

The initial data, $\gamma_{ij}$ and $K_{ij}$, evolve in time according to the evolution equations \begin{eqnarray}
\partial_t \gamma_{ij} & = & - 2 N K_{ij} + \bar{\nabla}_i \beta_j + \bar{\nabla}_j \beta_i, \\
\partial_t K_{ij}      & = & - \bar{\nabla}_i \bar{\nabla}_j N + N \left( \bar{R}_{ij} - 2 K_{ik} K^k_j + K K_{ij} \right) \nonumber \\
& & + \beta^k \bar{\nabla}_k K_{ij} + K_{ik} \bar{\nabla}_j \beta^k + K_{kj} \bar{\nabla}_i \beta^k. \label{eqn:KijEvol}
\end{eqnarray}

We use the conformal thin-sandwich decomposition of the initial-value equations \cite{york98}. The spatial metric $\gamma_{ij}$ is related to a background (or conformal) spatial metric $\tilde{\gamma}_{ij}$ \cite{lich44,york71,york72}, \begin{equation}
\gamma_{ij} = \psi^4 \tilde{\gamma}_{ij}.
\end{equation} The extrinsic curvature is split into its trace and trace-free parts, \begin{equation}
K_{ij} = A_{ij} + \frac{1}{3} \gamma_{ij} K.
\end{equation} The trace of the extrinsic curvature $K$ is the same in both the physical and conformal space, $K = \tilde{K}$. We provide a conformal weighting for the lapse function, $N = \psi^6 \tilde{N}$, and introduce the time derivative of the conformal metric, \begin{equation}
\tilde{u}_{ij} = \partial_t \tilde{\gamma}_{ij},
\end{equation} which transforms as $u_{ij} = \psi^4 \tilde{u}_{ij}$. 

In this decomposition the conformal extrinsic curvature is given by \begin{equation}
\tilde{A}_{ij} = \frac{1}{2\tilde{N}} \left[ ( \tilde{\mathbb L} \beta )_{ij} - \tilde{u}_{ij} \right], \label{eqn:fullAij}
\end{equation} and its conformal weighting $A_{ij} = \psi^{-2} \tilde{A}_{ij}$ and $A^{ij} = \psi^{-10} \tilde{A}^{ij}$ follows. The resulting conformal thin-sandwich decomposition of the initial-value equations is, in vacuum, \begin{eqnarray}
 \tilde{\nabla}^2 \psi & = & \frac{1}{12}\psi^5 K^2 - \frac{1}{8} \psi^{-7} \tilde{A}_{ij} \tilde{A}^{ij} \nonumber \\
& & \ \ \ \ \ + \frac{1}{8} \psi \tilde{R},  \label{eqn:CTSHCA} \\
\tilde{\nabla}_j \left( \frac{1}{2\tilde{N}} ({\mathbb L} \beta)^{ij} \right) & = & \frac{2}{3} \psi^6 \tilde{\nabla}^i K + \tilde{\nabla}_j \left( \frac{1}{2 \tilde{N}} \tilde{u}^{ij} \right) , \label{eqn:CTSMCA} \\
 \tilde{\nabla}^2 (\psi^7 \tilde{N} ) & = & \tilde{N} \psi^7 \left( \frac{7}{8} \psi^{-8} \tilde{A}_{ij} \tilde{A}^{ij} + \frac{5}{12} \psi^4 K^2 \right) \nonumber \\
& & \ \ \ \ \ + \psi^5 \left( \beta^l \tilde{\nabla}_l K - \partial_t K \right) , \ \label{eqn:CTSconstKA}
\end{eqnarray} where \begin{eqnarray}
( \tilde{\mathbb L} \beta)^{ij} & = & \tilde{\nabla}^i \beta^j + \tilde{\nabla}^j \beta^i - \frac{2}{3} \tilde{\gamma}^{ij} \tilde{\nabla}_k \beta^k, \\
\tilde{\Delta}_{\mathbb L} \beta^i & = & \tilde{\nabla}^2 \beta^i + \frac{1}{3} \tilde{\nabla}^i \tilde{\nabla}_j \beta^j + \tilde{R}^i_j \beta^j.
\end{eqnarray} Equation (\ref{eqn:CTSHCA}) is the CTS version of the Hamiltonian constraint (\ref{eqn:HC}). The momentum constraint (\ref{eqn:MC}) is now written as (\ref{eqn:CTSMCA}). Equation (\ref{eqn:CTSconstKA}) follows from specifying $\partial_t K$ in the ADM evolution equation for $K$, obtained from the trace of (\ref{eqn:KijEvol}). This equation provides us with a lapse function, $N = \psi^6 \tilde{N}$, and is necessary to complete the CTS system \cite{york03}. 

In this decomposition the free data are the conformal metric $\tilde{\gamma}_{ij}$ and trace of the extrinsic curvature $K$, and their time derivatives on the initial slice, $\tilde{u}_{ij}$ and $\partial_t K$. Throughout this paper we make the choices of conformal flatness, $\tilde{\gamma}_{ij} = f_{ij}$ (the flat metric in Cartesian coordinates), maximal slicing, $K = 0$, and initial stationarity of $K$ and the conformal spatial metric, $\tilde{u}_{ij} = \partial_t K = 0$. These stationarity choices are useful in the context of binary black holes in quasi-circular orbits \cite{mdhPhD,cook04}. With these choices, the conformal thin-sandwich equations become \begin{eqnarray} 
 \tilde{\nabla}^2 \psi &    =  &    - \frac{1}{8} \psi^{-7} \tilde{A}_{ij} \tilde{A}^{ij}, \  \label{eqn:CTSHC} \\
 \tilde{\Delta}_{\mathbb L} \beta^i - ( {\tilde{\mathbb L}} \beta )^{ij} \tilde{\nabla}_j \ln \tilde{N} &     =     & 0, \  \label{eqn:CTSMC} \\
 \tilde{\nabla}^2 (\psi^7 \tilde{N} ) & =  & \tilde{N} \psi^7 \left[ \frac{7}{8} \psi^{-8} \tilde{A}_{ij} \tilde{A}^{ij} \right]. \ \label{eqn:CTSconstK}
\end{eqnarray}  

Equation (\ref{eqn:CTSconstK}) is the maximal-slicing equation. It is a condition that, if the current slice is maximal, and we evolve to the next slice with a lapse that obeys (\ref{eqn:CTSconstK}), then that slice will also be maximal.

With $\tilde{u}_{ij} = 0$, the conformal extrinsic curvature (\ref{eqn:fullAij}) is now given by \begin{equation}
\tilde{A}_{ij} = \frac{1}{2\tilde{N}} ( \tilde{\mathbb L} \beta )_{ij}. \label{eqn:Aij}
\end{equation}

\section{The puncture method}
\label{sec:ctsp}

We wish to solve the equations (\ref{eqn:CTSHC}) -- (\ref{eqn:CTSconstK}) using the puncture approach \cite{brandt97}. In this approach the conformal factor for $n$ black holes is written as \begin{equation}
\psi = 1 + \sum_{i}^{n} \frac{m_i}{2r_i} + u. \label{eqn:psi}
\end{equation} Here $m_i$ are parameters that characterize the mass of each black hole. In the special case of a single Schwarzschild black hole, $m_1$ equals the ADM mass of the black hole. The $r_i$ are the distances of each black hole from the origin. Equation (\ref{eqn:psi}) gives the Brill-Lindquist conformal factor for $n$ momentarily stationary black holes \cite{brill63,misner57}, plus a function $u$, which describes the deviation in the conformal factor due to a non-zero extrinsic curvature. The function $u$ is regular over all of the conformal space \cite{brandt97}. The topology of this solution consists of one hypersurface that contains $n$ black holes, and each black hole connects through an Einstein-Rosen bridge to an additional hypersurface, giving a total of $n+1$ hypersurfaces connected by $n$ Einstein-Rosen bridges.

In the CTSP approach \cite{hannam03,tichy03} the conformal lapse $\tilde{N}$ is also split into an analytic singular and unknown regular part, \begin{equation}
\tilde{N} \psi^7 = N \psi = 1 + \sum_i^n \frac{c_i}{2 r_i} + v. \label{eqn:lapsesplit}
\end{equation} We are free to choose the constants $c_i$; they determine the value of the lapse at spatial infinity on the other hypersurfaces. However, negative choices of $c_i$ will lead to a lapse function that goes through zero, and this will cause division-by-zero problems in a numerical construction of $\tilde{A}_{ij}$ via (\ref{eqn:Aij}). For this reason, we choose positive values of $c_i$ in this paper. We investigate the effect of different choices of $c_i$ on the physical results in section \ref{sec:ctsp_BH}.

With these choices, the conformal thin-sandwich puncture (CTSP) system is \begin{eqnarray}
\tilde{\nabla}^2 u & = & - \frac{1}{8} \psi^{-7} \tilde{A}_{ij} \tilde{A}^{ij}, \label{eqn:CTSPHC} \\
\tilde{\nabla}^2 v & = & \tilde{N} \psi^7 \left[ \frac{7}{8} \psi^{-8} \tilde{A}_{ij} \tilde{A}^{ij}  \right], \label{eqn:CTSPconstK} \\
\tilde{\Delta}_{\mathbb L} \beta^i - ( {\tilde{\mathbb L}} \beta )^{ij} \tilde{\nabla}_j \ln \tilde{N} & = & 0.  \label{eqn:CTSPMC} 
\end{eqnarray} Brandt and Br\"{u}gmann \cite{brandt97} showed that the solution $u$ of (\ref{eqn:CTSPHC}) will be $C^2$ everywhere on $R^3$ if the conformal extrinsic curvature $\tilde{A}_{ij}$ diverges no faster that $1/r_i^3$ at each puncture. The same will be true for (\ref{eqn:CTSPconstK}) \cite{hannam03}. In the construction of $\tilde{A}_{ij}$ via (\ref{eqn:Aij}), the conformal lapse $\tilde{N}$ goes to zero as $r_i^6$. If $\tilde{A}_{ij}$ is to satisfy the divergence requirements of (\ref{eqn:CTSPHC}) and (\ref{eqn:CTSPconstK}), then $(\tilde{\mathbb L} \beta)_{ij}$ must go to zero at least as fast as $r_i^3$. If this requirement is met, then the CTSP equations will be regular over all of $R^3$, eliminating the need for any region of the computational domain to be excised. 

The particular solution that we obtain will depend on the boundary conditions we apply, both at the outer boundary and at the puncture(s). There is no need for puncture boundary conditions on equations (\ref{eqn:CTSPHC}) and (\ref{eqn:CTSPconstK}) if their source terms are smooth at the punctures. On the other hand, puncture conditions are needed for the momentum constraint (\ref{eqn:CTSPMC}), if we wish to avoid a trivial solution. For example, consider a single boosted black hole. We expect the shift vector to fall off to zero at spatial infinity. However, if we impose the outer boundary condition $\beta^i = 0$ on the shift vector, and {\it do not} place any condition on the shift vector at the puncture, a numerical code will yield the trivial solution $\beta^i = 0$ everywhere. A zero shift vector will lead to a zero extrinsic curvature, and will not represent a boosted black hole. 

In this work we will show that a black hole with non-zero linear momentum can be generated by imposing a value on the shift vector {\it at the puncture}, in addition to outer boundary conditions. The full CTSP system is solved at every point on a Cartesian grid, except at the punctures, where the momentum constraint is {\it not} solved, and instead at the $a$th puncture we require that \begin{equation}
\beta^i_a = A^i_a.
\end{equation} The values $A^i_a$ parametrize the linear momentum of the black hole, which will point in the direction of the vector $\hat{A}^i_a = A^i_a / |A|$ for each hole. In this prescription the shift vector will be zero at spatial infinity. We will provide a motivation for this procedure in the next section, where we consider solutions for a single black hole with linear momentum.

\section{A single black hole with linear momentum}
\label{sec:1BH}


Laguna \cite{laguna04} has found analytic perturbative solutions of the CTSP system (\ref{eqn:CTSPHC}) -- (\ref{eqn:CTSPMC}) that describe a single boosted or spinning black hole. Laguna's solutions yield the Bowen-York extrinsic curvature. In the case of a boosted black hole, these solutions are perturbative to second order in the momentum for $u$ and $v$, and linear for $\beta^i$. We will focus on the momentum constraint, which takes the form Laguna considered if we choose $u = v = 0$ (or $u = v = 1$ in Laguna's notation). The conformal factor and lapse splittings (\ref{eqn:psi}) and (\ref{eqn:lapsesplit}) are then \begin{eqnarray}
\psi & = & 1 + \frac{m}{2r}, \\
\tilde{N}\psi^7 & = & 1 + \frac{c}{2r}.
\end{eqnarray} The lapse parameter $c$ corresponds to the parameter $-b$ in \cite{laguna04}. The solution of the CTSP momentum constraint for a Bowen-York black hole with linear momentum $P^i = (P,0,0)$, located at the origin in Cartesian co-ordinates, is 
\begin{eqnarray}
\beta^x & = & \frac{-4 x^2 f(r,m,c) - g(r,m,c)}{20(m +2r)^6} P, \label{eqn:LagBetax} \\
\beta^y & = & - \frac{xy f(r,m,c)}{5(m + 2r)^6} P,  \label{eqn:LagBetay} \\
\beta^z & = & - \frac{xz f(r,m,c)}{5(m + 2r)^6} P, \label{eqn:LagBetaz}
\end{eqnarray} with \begin{eqnarray}
f(r,m,c) & = &  (m+c) m^2 + 12(m+c)m r \nonumber \\
& & \ \ \ \ \ + 60 (m+c)         r^2 + 160               r^3, \\
g(r,m,c) & = & 5(5m+c) m^4 + 60(5m+c)m^3 r \nonumber \\ 
& & \ \ \ \ \ +    2(749m+149c) m^2 r^2 \nonumber \\
& & \ \ \ \ \ + 8   (497m+97c) m r^3  \nonumber \\
& & \ \ \ \ \ + 120(49m+9c) r^4 + 4480r^5.
\end{eqnarray}

A calculation of the conformal extrinsic curvature via (\ref{eqn:Aij}) yields \begin{equation}
\tilde{A}^{ij} = \frac{1}{r^2} \left[ P^i n^j + P^j n^i + (f^{ij} - n^i n^j) P^k n_k \right], \label{eqn:bowenyorkaij}
\end{equation} where $n^i = x^i / r$ are normal vectors directed away from the puncture. Equation (\ref{eqn:bowenyorkaij}) is the Bowen-York extrinsic curvature for a black hole with linear momentum $P^i$ \cite{bowen80}. 

As $r \rightarrow \infty$, the conformal lapse approaches unity, $\tilde{N} \rightarrow 1$, and the momentum constraint (\ref{eqn:CTSPMC}) approaches its conformal transverse-traceless (CTT) decomposition form, \begin{equation}
\tilde{\Delta}_{\mathbb L} W^i = 0.
\end{equation} The solution $W^i$ that corresponds to a boosted Bowen-York black hole with momentum $P^i$ is \begin{equation}
W^i = - \frac{1}{4r} \left( 7 P^i + n^i n_j P^j \right). \label{eqn:BYvecpot}
\end{equation} The asymptotic form of (\ref{eqn:LagBetax}) -- (\ref{eqn:LagBetaz}) is \begin{equation}
\beta^i = 2W^i.
\end{equation}

The solution (\ref{eqn:LagBetax}) -- (\ref{eqn:LagBetaz}) is not unique: $\beta^i + V^i$ is also a solution of the momentum constraint, for any constant vector $V^i$. This choice will determine the value of the shift vector both at the puncture and asymptotically. The value of the shift vector at the puncture, $\beta^i_0$, is \begin{equation}
\beta^i_0 = - \frac{5m+c}{4m^2} P^i + V^i,
\end{equation} and its asymptotic form is \begin{equation}
\left. \beta^i \right|_{r \rightarrow \infty} = 2W^i + V^i.
\end{equation} In the solution presented in (\ref{eqn:LagBetax}) -- (\ref{eqn:LagBetaz}), we have chosen $V^i = 0$, and the shift vector is $\beta^i = [-(5m+c)P/(4m^2),0,0]$ at the puncture, and asymptotically approaches twice the Bowen-York vector potential for a boosted black hole. Laguna's form of the solution is zero at the puncture, and has the asymptotic form $\beta^i = (5m+c)P^i/(4m^2) + 2 W^i$. 

This solution motivates the condition on the shift vector at the puncture that we introduced in section \ref{sec:ctsp}. To numerically construct a single black hole with linear momentum, one can solve (\ref{eqn:CTSPHC}) -- (\ref{eqn:CTSPMC}), requiring that, at the puncture, \begin{equation}
\beta^i_0 = - \frac{(5m+c)}{4m^2} P^i , \label{eqn:beta_punc}
\end{equation} where $c > 0$, and apply an outer boundary condition consistent with the Bowen-York vector potential (\ref{eqn:BYvecpot}). We will show in section \ref{sec:cvg} that this technique does indeed produce a black hole with linear momentum, although the magnitude of the linear momentum will be equal to $P$ only when we make the assumption that $u = v = 0$ in the solution of the momentum constraint. Generalization to two (or more) black holes is straightforward.

Given the CTSP system of equations (\ref{eqn:CTSPHC}) -- (\ref{eqn:CTSPMC}), we must choose appropriate outer boundary conditions. Requiring that our solutions be asymptotically flat implies that we can apply $1/r$ Robin boundary conditions to the functions $u$ and $v$. We are therefore assuming that beyond the outer boundary of the numerical grid the functions $u$ and $v$ behave as, \begin{eqnarray}
u & = & \frac{k_1}{r}, \label{eqn:ubc} \\
v & = & \frac{k_2}{r}, \label{eqn:vbc}
\end{eqnarray} where $k_1$ and $k_2$ are constants. Note that in a numerical code the finite outer boundary will introduce errors, and as a result the values of $k_1$ and $k_2$ may vary across the surface of the outer boundary. To impose $1/r$ fall-off on a function $f$, we impose \begin{equation}
N^i \partial_i (rf) = 0, \label{eqn:scalar_Robin}
\end{equation} where $N^i$ is the unit normal to the outer boundary. The outer boundary condition (\ref{eqn:scalar_Robin}) is applied to the solutions of both the Hamiltonian constraint (\ref{eqn:CTSPHC}) and the maximal-slicing equation (\ref{eqn:CTSPconstK}).

In this paper we will use three types of outer boundary condition on the shift vector. In Section \ref{sec:cvg} we will use a Dirichlet outer boundary condition and prescribe a value for the shift vector at each point on the outer boundary, \begin{equation}
\beta^i = B^i, \label{eqn:dirichlet}
\end{equation} where $B^i$ is a known vector. In the case where we choose $u = v = 0$, $B^i$ is the analytic Laguna solution (\ref{eqn:LagBetax}) -- (\ref{eqn:LagBetaz}). In the general CTSP case we could choose $B^i = 0$, but this would introduce an outer boundary error that falls off only linearly as the outer boundary is moved out. 

The second type of outer boundary condition we use is a scalar Robin boundary condition, as described above for the solutions of the Hamiltonian constraint and maximal-slicing equation, and given by (\ref{eqn:scalar_Robin}). The error in this boundary condition will fall off quadratically as the outer boundary is moved out.

A third type of outer boundary condition makes use of our prior knowledge of the asymptotic angular dependence of the shift vector. As we pointed out earlier, the CTS momentum constraint approaches the form of the CTT momentum constraint for large $r$. The asymptotic behavior of any solution of the CTT momentum constraint with symmetries consistent with a boosted black hole \cite{omurch92} suggests that the shift vector will have the asymptotic form \begin{equation}
\beta^i = \frac{k_3}{r} \left( 7 \hat{P}^i + \hat{P}^k n_k n^i \right), \label{eqn:shiftbc}
\end{equation} where $\hat{P}^i$ is a unit vector in the direction of the linear momentum. In the case of a black hole boosted in the $x$-direction, we choose $\hat{P}^i = (1,0,0)$. The resulting Robin outer boundary condition is \begin{equation}
N^j \partial_j ( r \beta^i) = \frac{k_3}{r} \left[ (N^i - 2 n^i N^j n_j) P^k n_k + n^i (P_j N^j) \right]. \label{eqn:vector_Robin}
\end{equation}

Note that this is an approximate vector Robin boundary condition since we are assuming a direction for the momentum $P^i$ in using (\ref{eqn:shiftbc}) to determine $k_3$ separately for each component of the shift vector. 

All three of the shift outer boundary conditions (\ref{eqn:scalar_Robin}), (\ref{eqn:dirichlet}) and (\ref{eqn:vector_Robin}) are investigated in Sections \ref{sec:cvg} and \ref{sec:ctsp_BH}.

\section{Numerical techniques and convergence tests}
\label{sec:cvg}

We solve the CTSP system (\ref{eqn:CTSPHC}) -- (\ref{eqn:CTSPMC}) with a multigrid elliptic solver, which is a modification of the {\tt BAM\_Elliptic} solver in the Cactus infrastructure \cite{cactus}. The CTSP equations are coded with a second-order finite-differencing stencil as a coupled elliptic system on a Cartesian grid.

In all solutions presented here the code imposes a puncture condition of the form (\ref{eqn:beta_punc}). This is done by placing the black-hole puncture at a grid point on the numerical grid. All of the CTSP variables are well-behaved at the puncture, so there is no problem placing the puncture on a grid point. We require (\ref{eqn:beta_punc}) at the puncture, and do not solve the momentum constraint there. The remaining two CTSP equations, (\ref{eqn:CTSPHC}) and (\ref{eqn:CTSPconstK}), are solved at all points on the numerical grid. 

\begin{figure}[!ht]
\begin{center}
\includegraphics[scale=1.21]{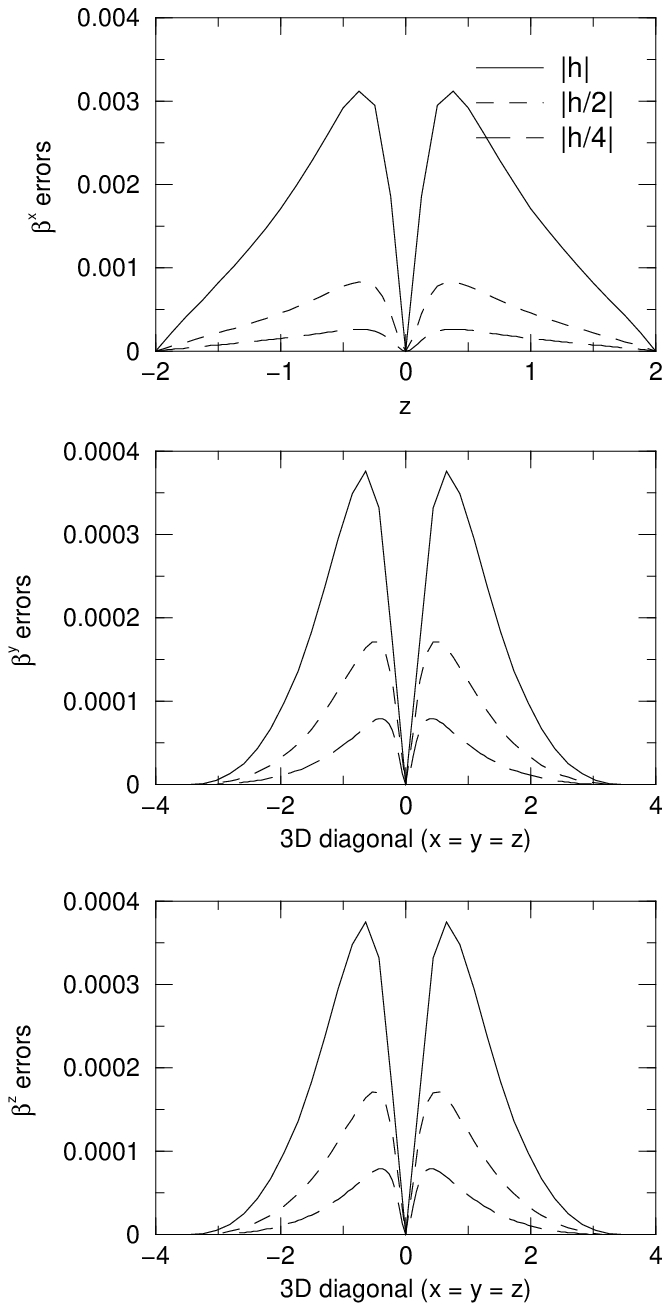}
\caption{Convergence behavior of the errors in the momentum constraint solution along the $z$-axis ($\beta^x$) and the 3D diagonal ($\beta^y$ and $\beta^z$), with outer boundary at 2$m$.}
\label{fig:LagCvg2.0}
\end{center}
\end{figure}

\begin{figure}[!ht]
\begin{center}
\includegraphics[scale=1.12]{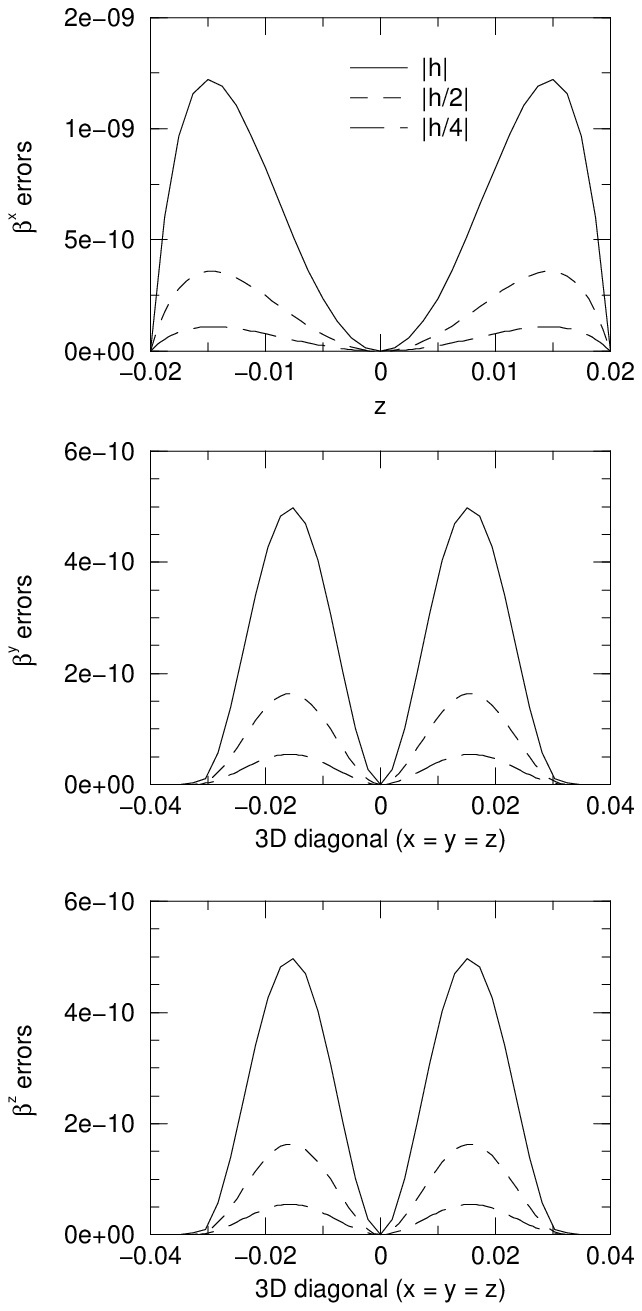}
\caption{Convergence behavior of the momentum constraint along the $z$-axis ($\beta^x$) and the 3D diagonal ($\beta^y$ and $\beta^z$), with outer boundary at 0.02$m$.}
\label{fig:LagCvg0.02}
\end{center}
\end{figure}

As a first illustration of these techniques, we consider the Laguna solution, i.e., we choose $u = v = 0$ and solve only the momentum constraint. We employ Dirichlet outer boundary conditions, imposing values on the shift vector at the outer boundary equal to the solution (\ref{eqn:LagBetax}) -- (\ref{eqn:LagBetaz}), making the choice $m = 1$, $c = 1$ in the decompositions of $\psi$ and $\tilde{N} \psi^7$ in (\ref{eqn:psi}) and (\ref{eqn:lapsesplit}). At the puncture we impose the condition (\ref{eqn:beta_punc}). We solve the momentum constraint for a single black hole with linear momentum in the $x$-direction of magnitude $P = 0.5m$. Convergence plots for this case are shown in Figures \ref{fig:LagCvg2.0} and \ref{fig:LagCvg0.02}, which represent solutions with outer boundaries of 2.0$m$ and 0.02$m$ respectively. These outer boundaries would be far too close for a general problem with Robin outer boundary conditions on the shift vector, but are possible in this case, where the analytic outer boundary values are known. All solutions were found with grid sizes of $32^3$, $64^3$ and $128^3$ points. The solution with outer boundary at $2m$ therefore has a base resolution of $h = 0.125m$. 

The figures show the disagreement between $\beta^i$ and the solution (\ref{eqn:LagBetax}) -- (\ref{eqn:LagBetaz}) for three grid resolutions. If the solutions were second-order convergent, the errors would fall by a factor of four each time the grid resolution $h$ was halved. First-order convergence would cause the errors to fall by a factor of two. 

The solutions in Figure \ref{fig:LagCvg2.0} show worse than second-order convergence, but better than first-order convergence. We found that the convergence behavior slowly improved as finer resolutions were used, but even at the extremely fine resolution of $h = 1.25 \times 10^{-3}m$ (the base resolution in Figure \ref{fig:LagCvg0.02}), we do not see full second-order convergence. 

The poor convergence behavior appears to be due to the use of the puncture condition (\ref{eqn:beta_punc}). This condition effectively applies an inner boundary condition at a single point, unlike a true inner boundary condition, in which a condition is applied to an entire inner surface. 

\begin{figure}[!ht]  
\begin{center}
\includegraphics[scale=0.45]{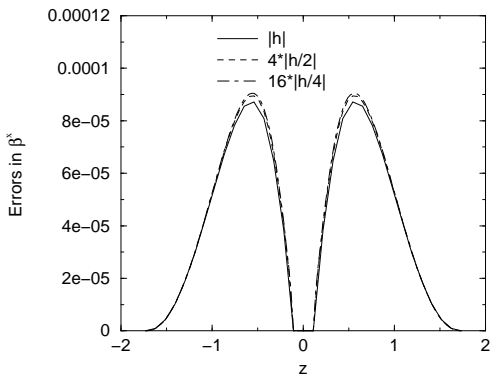}
\caption{Convergence of $\beta^x$ (along $z$-axis) when an inner boundary condition is used, and the outer boundary is at 1.0$m$. Differences were scaled assuming second-order convergence.}
\label{fig:Excision}
\end{center}
\end{figure}

The effect of the puncture can be seen by solving the system with a true inner boundary. A cubical region around the puncture was excised, and the analytic solution imposed on the resulting inner boundary. Figure \ref{fig:Excision} shows the convergence behavior of the $x$-component of the shift vector for this case, with an outer boundary at 1.0$m$ and a base resolution of $h = 0.0625m$. It is clear that in this case we have second-order convergence. The use of a puncture condition reduces the rate of convergence. This is unfortunate, but not catastrophic: we still have a convergent code, and since we know the correct condition to apply at the black-hole puncture, but {\it do not} know what inner boundary condition to apply on an arbitrary inner boundary, the use of a puncture condition remains advantageous. 

\section{A single boosted CTSP black hole}
\label{sec:ctsp_BH}

We are now ready to consider the full CTSP system, where the Hamiltonian constraint (\ref{eqn:CTSPHC}) and maximal-slicing equation (\ref{eqn:CTSPconstK}) are coupled to the momentum constraint (\ref{eqn:CTSPMC}). In order to save computer resources, we solve the full CTSP system on one octant of the numerical grid (i.e., for only positive $x$, $y$ and $z$) by employing the known symmetries of the solutions. In the case of a single black hole boosted in the $x$-direction, the CTSP variables exhibit the symmetries listed in Table \ref{tab:sym}. Table \ref{tab:sym} also shows the Robin outer boundary conditions that we use. We will apply both scalar and vector Robin outer boundary conditions to the shift vector, as described in Section \ref{sec:1BH}, and compare their effect on the solutions. 

\begin{table}
\begin{center}
\begin{tabular}{ccccc} 
          & $x = 0$ & $y = 0$ & $z = 0$ &  $r \rightarrow \infty$ \\ \hline
$u$       & even    &  even   &  even   &  $1/r$ \\
$v$       & even    &  even   &  even   &  $1/r$ \\
$\beta^x$ & even    &  even   &  even   &  $1/r$ or $(7 + x^2/r^2)/r$ \\
$\beta^y$ &  odd    &  odd    &  even   &  $1/r$ or $xy/r^3$ \\
$\beta^z$ &  odd    &  even   &  odd    &  $1/r$ or $xz/r^3$ \\
\end{tabular}
\caption{Coordinate plane symmetries and outer boundary conditions for CTSP variables for a single boosted black hole with momentum in the $x$-direction.} \label{tab:sym}
\end{center}
\end{table}

\begin{figure*}[!ht]  
\begin{center}
\includegraphics[scale=1.05]{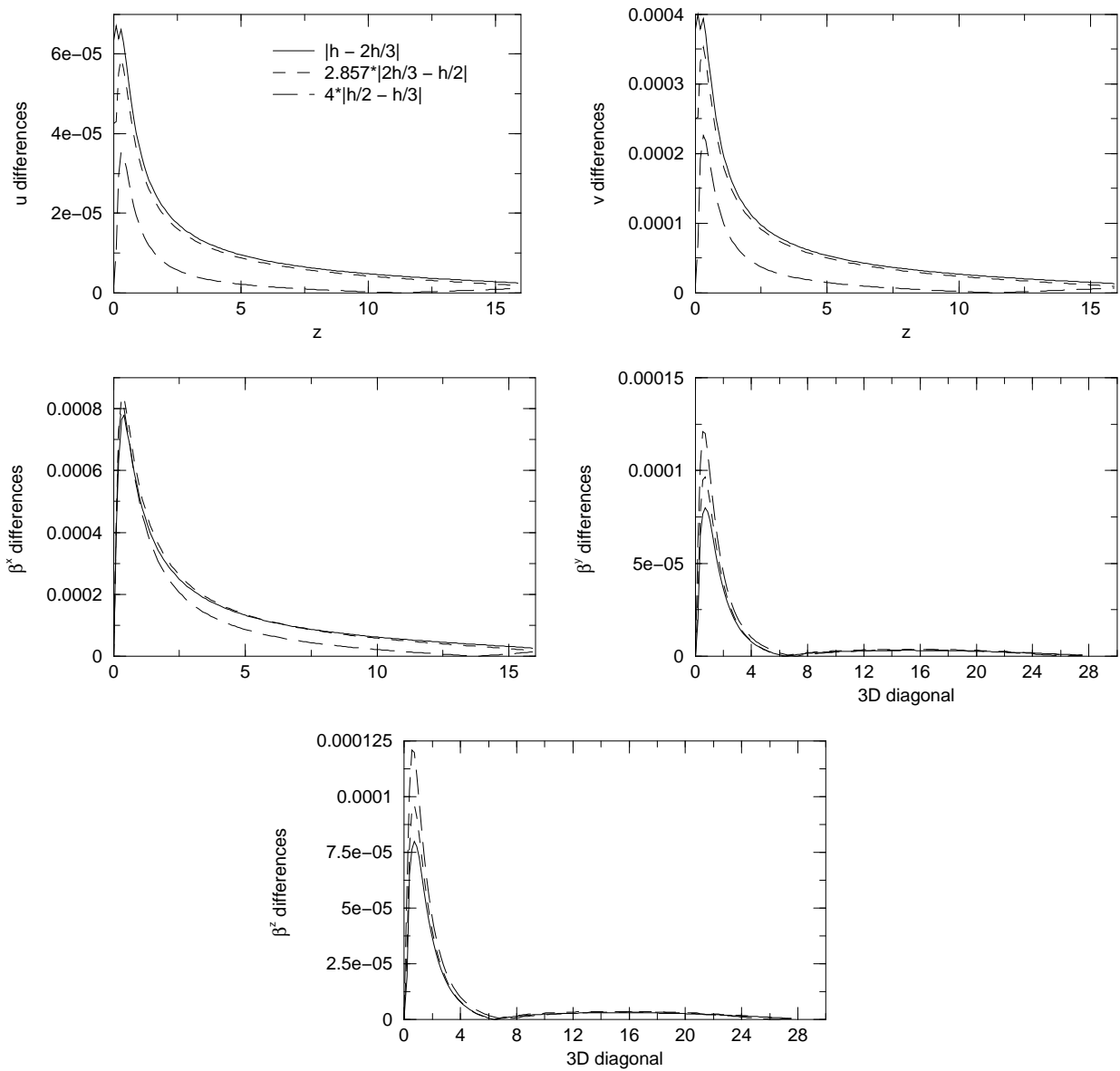}
\caption{Convergence of the full CTSP system for one boosted black hole with outer boundary at 16.0$m$, using scalar Robin outer boundary conditions on the shift vector. The differences are scaled assuming second-order convergence. The base grid spacing is $h = 0.125m$.}
\label{fig:cvg16r_1BH}
\end{center}
\end{figure*}

Figure \ref{fig:cvg16r_1BH} shows a convergence plot for a single black hole with the outer boundary at $16m$, and a base grid spacing of $h = 0.125m$. For these solutions a scalar Robin outer boundary condition (\ref{eqn:scalar_Robin}) was used for all CTSP variables, including the shift vector. The differences between solutions with resolutions $h$, $2h/3$, $h/2$ and $h/3$ are plotted, corresponding to grid sizes of $128^3$, $192^3$, $256^3$ and $384^3$ points. The differences are scaled assuming second-order convergence. These results are consistent with the test problem considered in section \ref{sec:cvg}: the solutions are between first- and second-order convergent. In particular, $\beta^y$ and $\beta^z$ show poor convergence near the puncture, and we can also see that convergence deteriorates a little near the outer boundary. 

\begin{figure*}[!ht]  
\begin{center}
\includegraphics[scale=1.02]{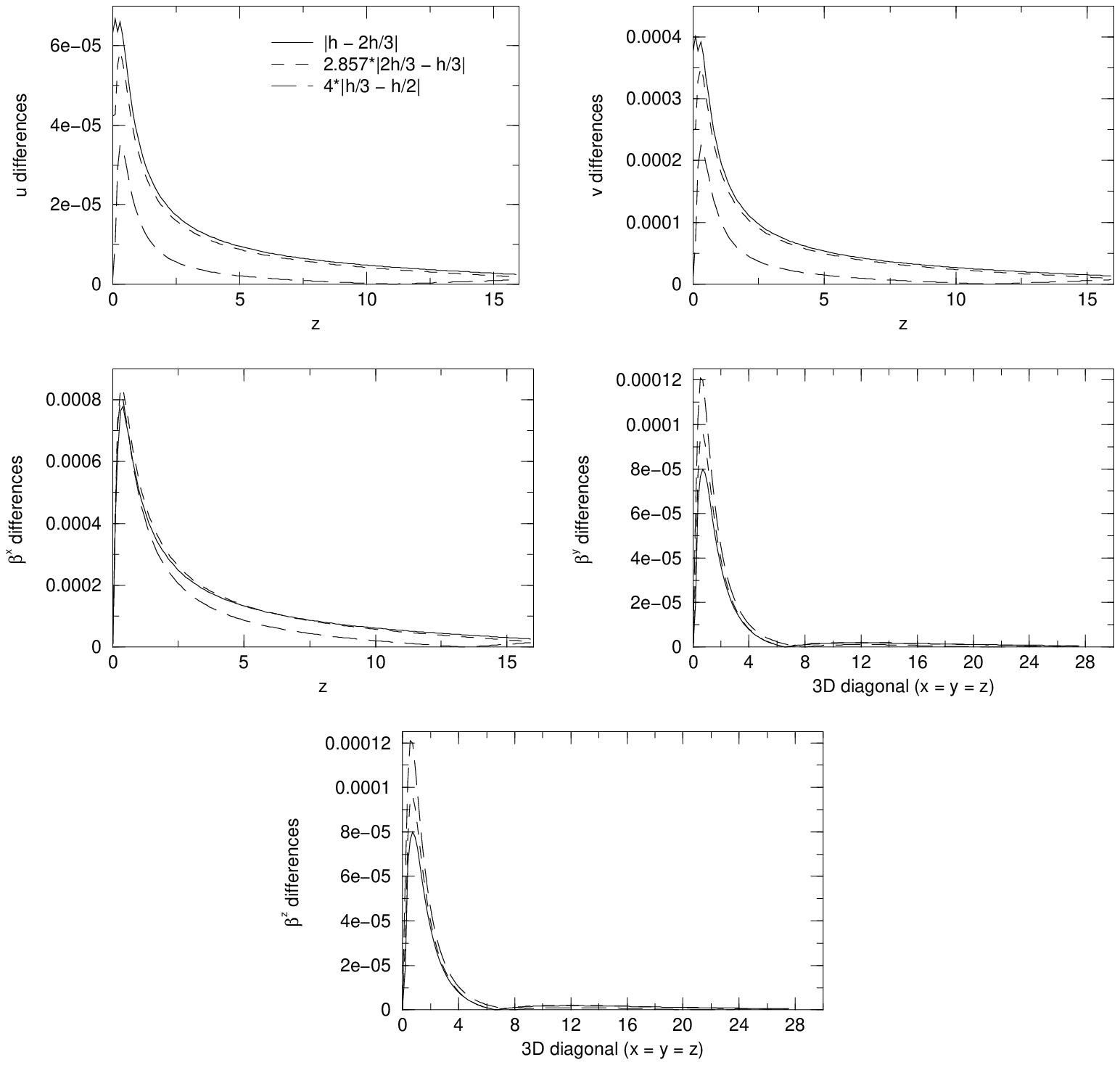}
\caption{Convergence of the full CTSP system for one boosted black hole with outer boundary at 16$m$, using vector Robin outer boundary conditions on the shift vector. The differences are scaled assuming second-order convergence. The base grid spacing is $h = 0.125m$.}
\label{fig:cvg16r_1BH_vec}
\end{center}
\end{figure*}

Figure \ref{fig:cvg16r_1BH_vec} shows convergence results for the same problem, but with the vector Robin outer boundary condition (\ref{eqn:vector_Robin}) applied to the shift vector. These plots show that both types of Robin outer boundary conditions have the same convergence properties.

\begin{figure*}[!ht]  
\begin{center}
\includegraphics[scale=1.01]{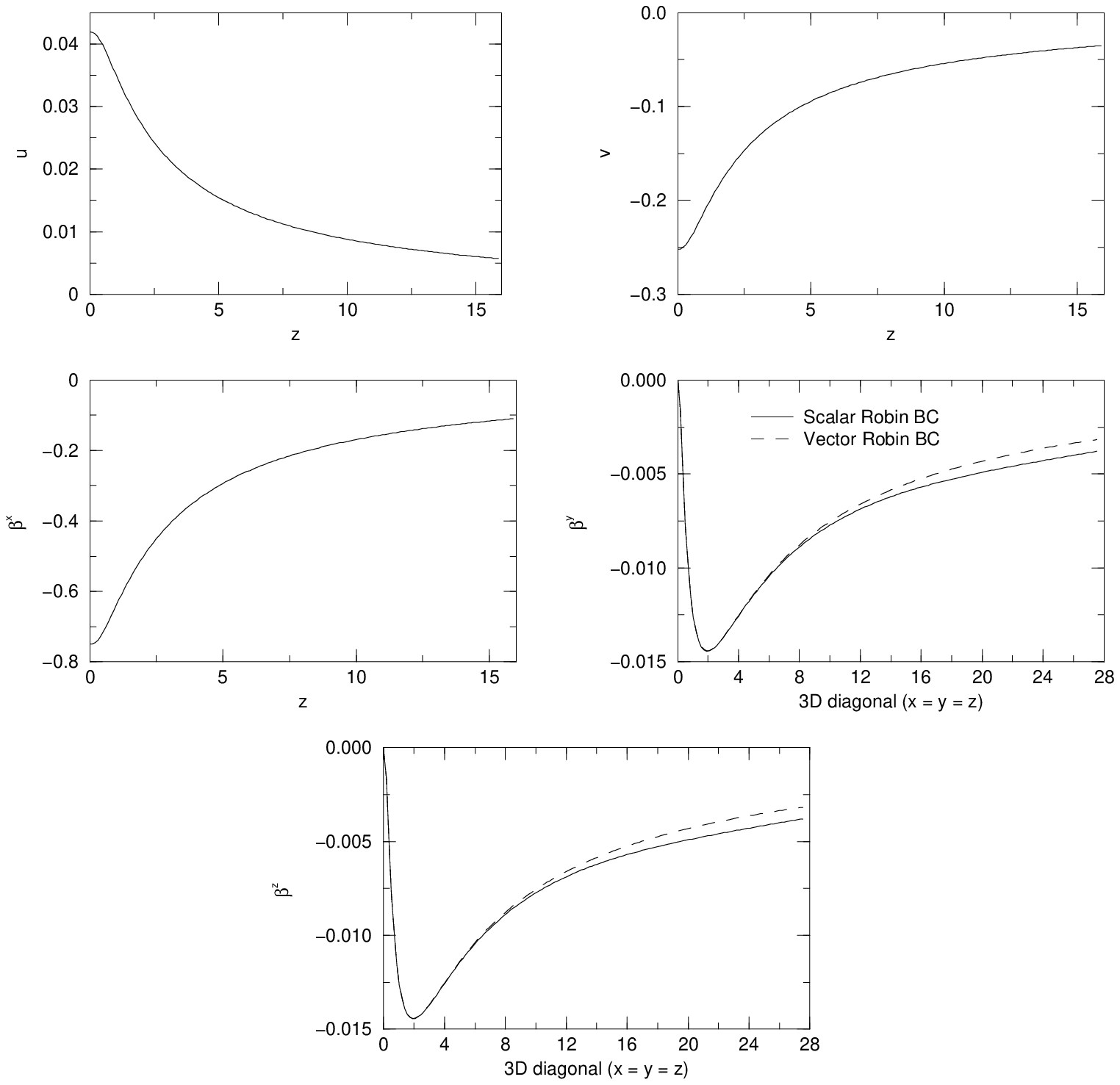}
\caption{Solutions of the full CTSP system for one boosted black hole with outer boundary at 16$m$, with $h = m/24$. The solutions for $u$, $v$ and $\beta^x$ did not differ substantially when either scalar or vector Robin outer boundary conditions were applied to the shift vector. The solutions of $\beta^y$ and $\beta^z$ are shown for both types of outer boundary condition.}
\label{fig:solns}
\end{center}
\end{figure*}

Figure \ref{fig:solns} shows the CTSP solutions for this problem, with $h = m/24$. The solutions of $u$, $v$ and $\beta^x$ are almost independent of whether we use scalar or vector Robin outer boundary conditions on the shift vector. However, the solutions of $\beta^y$ and $\beta^z$ differ significantly at the outer boundary, and the solutions for both choices of shift vector outer boundary condition are shown in Figure \ref{fig:solns}.  

Having generated solutions to the CTSP equations, we now calculate a number of physical quantities on the resulting spacetime slice.

In the analytic solution (\ref{eqn:LagBetax}) -- (\ref{eqn:LagBetaz}) we know the relationship between the value of the shift vector at the puncture and the linear momentum of the black hole. This relationship does not hold for the full CTSP system. We calculate the linear momentum of the initial data using \cite{bowen80,cook94} \begin{equation}
P^i = \frac{1}{8 \pi} \oint\tilde{A}_{ij} d^2 S_j. \label{eqn:pxint}
\end{equation} The surface integral is computed at the outer boundary of the computational grid. The integral was constructed using global Killing vectors of the conformal space (which are asymptotic Killing vectors of the physical space) and can in fact be computed at any radius, so long as the surface surrounds the puncture. For the setup we have described, the symmetries will enforce $P^y = P^z = 0$. 

\begin{figure}[!ht]  
\begin{center}
\includegraphics[scale=0.45]{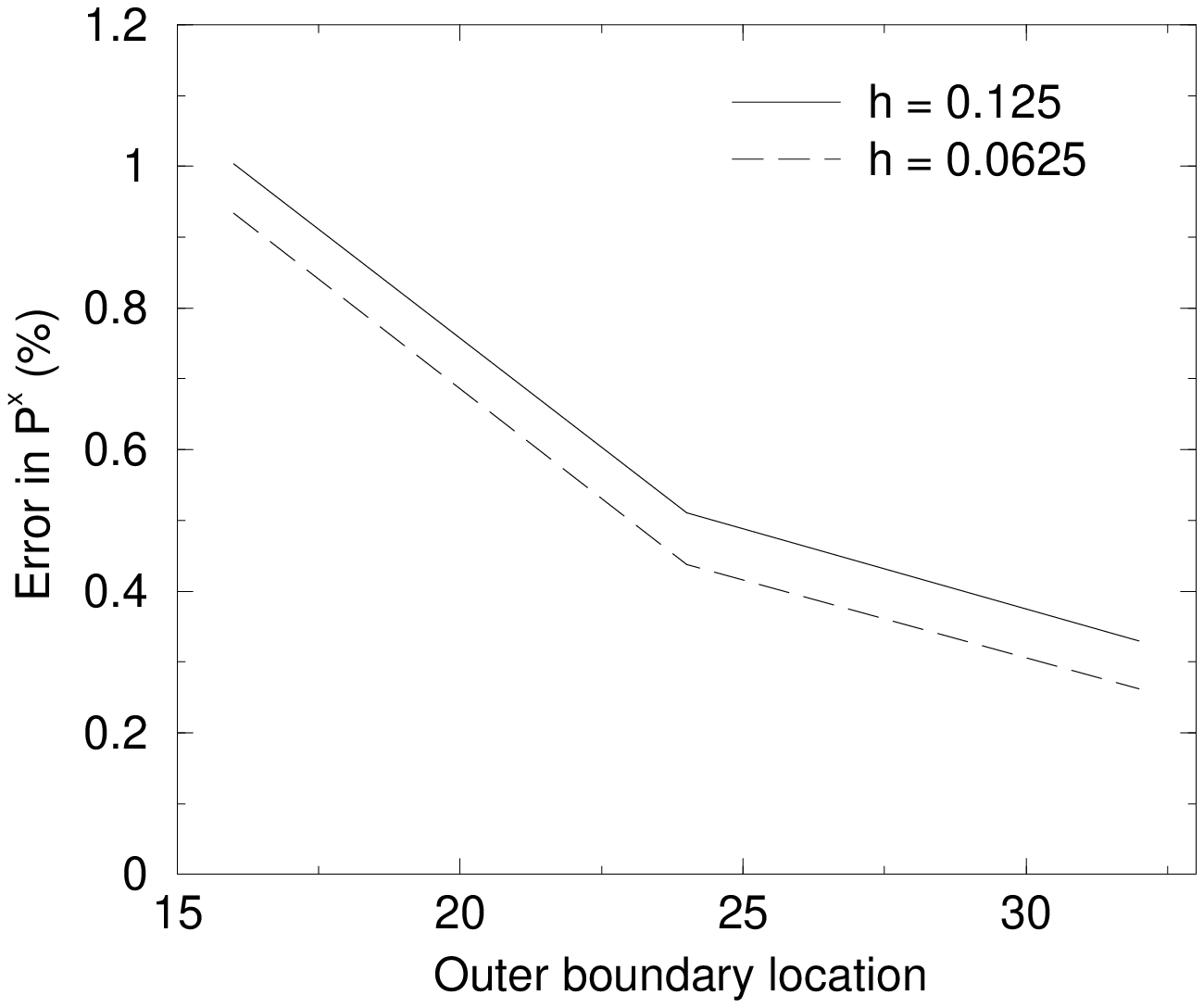}
\caption{The percentage error in $P^x$ for different choices of resolution and outer boundary for the Laguna case, with $c = m = 1$, $\beta^i_0 = -0.75$, and analytic $P^x = 0.5m$.}
\label{fig:Px_test}
\end{center}
\end{figure}

As stated above, the linear momentum of the solution is known analytically in the Laguna case, with (\ref{eqn:CTSPHC}) and (\ref{eqn:CTSPconstK}) solved using the Bowen-York extrinsic curvature in the source terms. This test case allows us to measure the accuracy of the solutions with respect to a physical quantity. Figure \ref{fig:Px_test} shows the error in the linear momentum for a single boosted black hole with $m = 1$, $c = 1$, $\beta^i_0 = (-0.75,0,0)$, and with the vector Robin boundary condition (\ref{eqn:vector_Robin}) applied to the shift vector. The analytic solution has $P^x = 0.5 m$. The code was run with two different resolutions $h = 0.125m$ and $h = 0.0625m$, and at three different choices of outer boundary location, $16m$, $24m$ and $32m$. We find that the location of the outer boundary has a far greater effect on the accuracy of the solutions than the resolution. As such, for subsequent runs we choose $h = 0.125m$ and an outer boundary at $32m$ as being an optimal combination of accuracy and economy of computational resources. 

\begin{figure}[!ht]  
\begin{center}
\includegraphics[scale=0.45]{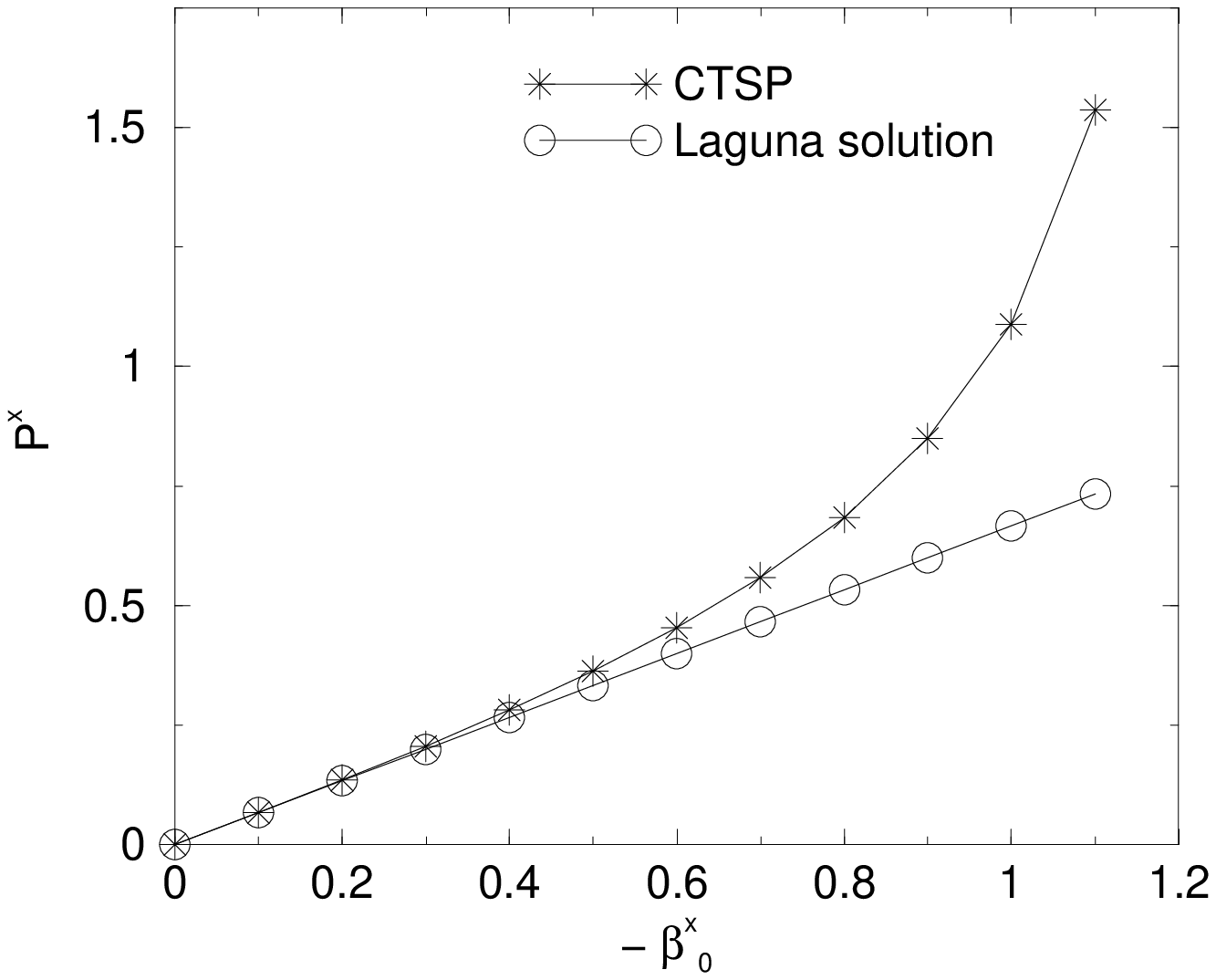}
\caption{The linear momentum $P_x$ of one CTSP black hole as a function of the value of $\beta^x_0$ at the puncture.}
\label{fig:Px}
\end{center}
\end{figure}

\begin{figure}[!ht]
\begin{center}
\includegraphics[scale=0.45]{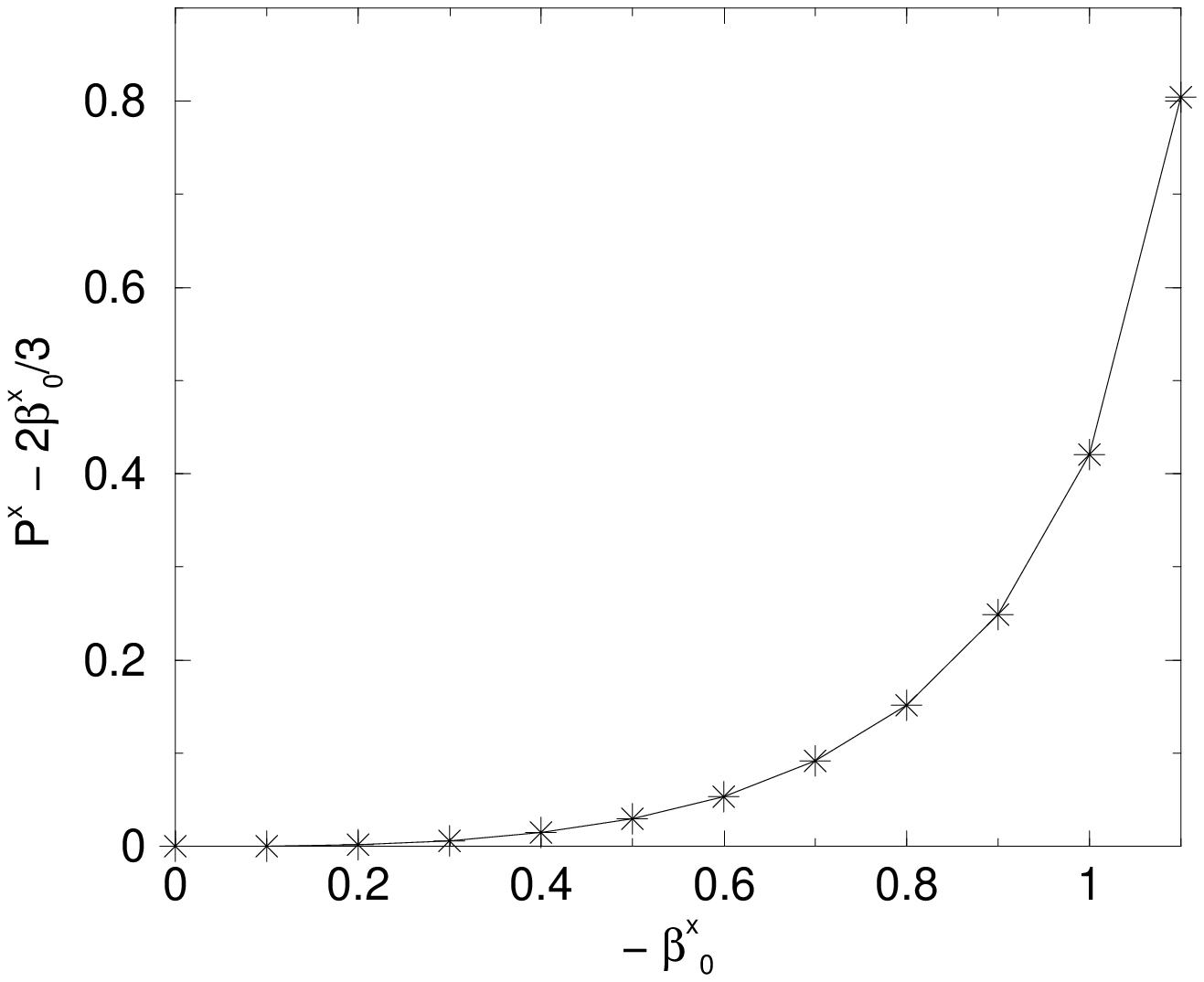}
\caption{The disagreement in linear momentum $P_x$ between the Laguna and full CTSP solutions for one black hole, as a function of the value of $\beta^x_0$ at the puncture.}
\label{fig:Px_diff}
\end{center}
\end{figure}

Figure \ref{fig:Px} shows the linear momentum in the $x$-direction as a function of the puncture value of $\beta^x_0$ for full CTSP solutions with vector Robin outer boundary conditions on the shift vector. Figure \ref{fig:Px_diff} shows how this value differs from that of the Laguna solution; as we would expect, the Laguna solution is a good approximation for small values of $\beta^x_0$. 

For a given initial-data set, we can also calculate the total ADM mass, $M_{ADM}$. In the conformal space, this is given by \cite{omurch74c} \begin{equation}
M_{ADM} = - \frac{1}{2\pi} \oint_{\infty} \tilde{\nabla}^i \psi d^2 S_i. \label{eqn:ADMintegral} 
\end{equation} In terms of the puncture splitting of the conformal factor (\ref{eqn:psi}) the ADM mass can be rewritten as a volume integral \cite{baum00} \begin{equation}
M_{ADM} = \sum_i^n m_i + \frac{1}{16\pi} \int \psi^{-7} \tilde{A}_{ij} \tilde{A}^{ij} dV.
\label{eqn:ADMintsplit}
\end{equation} 

The ADM mass integral is over all space, but the numerical grid covers only a finite region. In order to estimate the contribution to the integral (\ref{eqn:ADMintsplit}) from beyond the numerical grid, we need to estimate the values of the CTSP variables in that region. Following Baumgarte \cite{baum00} we do this by assuming that beyond the numerical grid the CTSP variables behave exactly as we required in their respective outer boundary conditions. For example, we required that \begin{equation}
u = \frac{k_1}{r},
\end{equation} at the outer boundary. Given some radial vector from the origin to a point on the outer boundary of the grid, we can calculate the constant $k_1$ at that point. The value of $u$ at any point along that line, beyond the grid, can now be estimated. The accuracy of $k_1$ depends on the location of the outer boundary and the resolution of the grid, but the calculation of $u$ beyond the grid is consistent with the outer boundary condition. 

A similar technique can be applied to the shift vector and $v$. However, an alternative approach is to recall that at the outer boundary the shift vector will behave like the Bowen-York vector potential for a single black hole, as described in the discussion before (\ref{eqn:shiftbc}). When the conformal extrinsic curvature is calculated using (\ref{eqn:Aij}) it will resemble the Bowen-York extrinsic curvature (\ref{eqn:bowenyorkaij}) at the outer boundary, up to a multiplicative factor. That factor will be related to the linear momentum of the black hole. Instead of calculating that factor numerically, we can simply use the linear momentum calculated with (\ref{eqn:pxint}) and construct the extrinsic curvature outside the numerical grid with (\ref{eqn:bowenyorkaij}) and use this in the ADM-mass integral (\ref{eqn:ADMintsplit}). The values of $\beta^i$ and $v$ outside the numerical grid are now unnecessary. The code we used to calculate the ADM mass integral was adapted from a code written by Miranda Dettwyler to perform the same calculation on Bowen-York puncture data \cite{mhdMS}. 

\begin{figure}[!ht]  
\begin{center}
\includegraphics[scale=0.45]{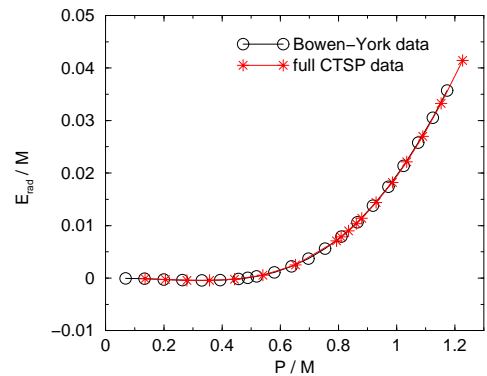}
\caption{Radiation content of single black-hole initial-data sets, with $m = c = 1$. Comparison of Laguna case with full CTSP data.}
\label{fig:rad}
\end{center}
\end{figure}

Given the ADM mass and total linear momentum of a data set, we can estimate the maximum possible amount of gravitational radiation that these initial-data sets contain. The maximum gravitational radiation content can be defined as \begin{equation}
\frac{E_{rad}}{M} = \frac{\sqrt{E_{ADM}^2 - P^2}}{M} - 1,
\end{equation} where $M$ is the puncture ADM mass of the black hole, defined as the ADM mass as calculated on the second hypersurface \cite{brandt97}. The puncture ADM mass is also often referred to as the bare mass of the black hole \cite{brill63,brandt97}. The puncture mass for a single black hole is given by \begin{equation}
M = m(1 + u_0),
\end{equation} where $u_0$ is the value of the function $u$ at the puncture and $m$ is the mass parameter in (\ref{eqn:psi}). Figure \ref{fig:rad} shows the maximum radiation content of a number of single black-hole initial-data sets, and a comparison with similar cases for Bowen-York puncture black holes. The Bowen-York puncture data were generated by solving the full CTSP system, but ignoring $u$ and $v$ in the construction of the conformal extrinsic curvature. This is similar to the Laguna case described in Section \ref{sec:cvg}, except that now the Hamiltonian constraint and maximal-slicing equations are also solved, and $u$ and $v$, although they are ignored in constructing the terms in the momentum constraint, are not zero. All solutions were found with an outer boundary of $32m$, with resolution $h = 0.125m$. Due to small systematic errors in the calculation of $E_{ADM}$ and $P$, the ADM mass is underestimated by about 0.1\%, and the linear momentum is overestimated by about 0.3\%. These errors cause the calculated radiation content to be too low. This is clearest for small values of $P/M$, for which the calculated maximum radiation content is negative. However, these are systematic errors, and do not prevent comparison between data sets generated and analyzed by the same procedure.   

Figure \ref{fig:rad} shows that single boosted CTSP and Bowen-York puncture black holes have the same maximum radiation content. They are indistinguishable from the point of view of the covariant quantities that we can calculate, the total ADM mass of the spacetime, the bare mass, and the linear momentum. 

\begin{figure}[!ht]  
\begin{center}
\includegraphics[scale=0.45]{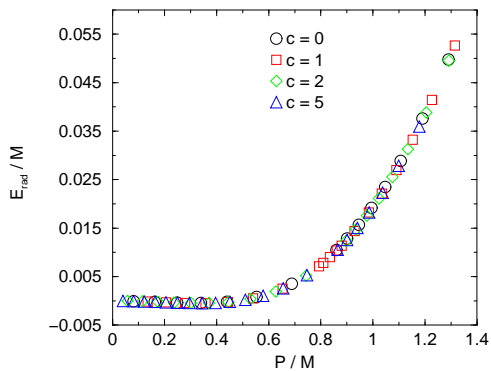}
\caption{Comparison of radiation content of single black hole initial-data sets, for various choices of the lapse parameter $c > 0$. The choice of $c$ has a minimal effect on the physical properties of the initial data.}
\label{fig:rad_c}
\end{center}
\end{figure}

There remains one free parameter in the CTSP approach, the constant $c$ in the lapse decomposition (\ref{eqn:lapsesplit}). We investigate the effect of different choices of the constant $c$ on CTSP boosted black-hole initial data in Figure \ref{fig:rad_c}. We look at the maximum radiation content of initial-data sets for $c = 0,1.0, 2.0$ and $5.0$. The only limitation on $c$ is that the lapse not become zero or negative at any point on the computational grid; if this is allowed to happen, the numerical code will encounter division-by-zero errors when calculating the extrinsic curvature via (\ref{eqn:Aij}). It is clear from Figure \ref{fig:rad_c} that the choice of $c$ has no affect on the maximum radiation content of the initial data.

\section{Conclusions}

We have completed the incorporation of the conformal thin-sandwich decomposition of the initial-value equations of general relativity into the puncture framework that was begun in \cite{hannam03}. We have developed a technique for constructing boosted black holes in the CTSP framework by specifying a value for the shift vector at the punctures, and have numerically constructed single black-hole initial-data sets using this prescription. 

We have studied the maximum radiation content of boosted CTSP black holes, and found that it equals the maximum radiation content of single boosted Bowen-York black holes. This result suggests that single boosted CTSP black holes are nearly indistinguishable from single boosted Bowen-York black holes. It would be necessary to numerically evolve both types of initial data to verify this conjecture. These results are independent of the free slicing parameter, $c$.

In the binary black-hole case, we do not expect CTSP and Bowen-York initial-data sets to be equivalent. Tichy and Br\"{u}gmann \cite{tichy04} have shown that binary Bowen-York black holes do not satisfy $\tilde{u}_{ij} = 0$, and therefore cannot be a solution of the CTSP equations with that quasiequilibrium choice. Similarly, there is no reason to expect that the physical properties of binary black-hole initial-data sets will be independent of the slicing parameters $c_i$. A detailed study of binary black-hole initial data will be made in a future publication.

\acknowledgments

M.H. would like to thank Carlos Lousto, Manuela Campanelli, Yosef Zlochower and Steven Lau for useful discussions. This work was supported in part by NASA grant No. NAG5-13396 to the Center for Gravitational Wave Astronomy at the University of Texas at Brownsville, NSF grants No. PHY-0140326 and PHY-0354867 to the University of Texas at Brownsville, and No. PHY-0140100 to Wake Forest University. Numerical results were obtained on the CGWA Funes cluster.

\bibliography{ctsp}

\begin{thebibliography}{33}
\expandafter\ifx\csname natexlab\endcsname\relax\def\natexlab#1{#1}\fi
\expandafter\ifx\csname bibnamefont\endcsname\relax
  \def\bibnamefont#1{#1}\fi
\expandafter\ifx\csname bibfnamefont\endcsname\relax
  \def\bibfnamefont#1{#1}\fi
\expandafter\ifx\csname citenamefont\endcsname\relax
  \def\citenamefont#1{#1}\fi
\expandafter\ifx\csname url\endcsname\relax
  \def\url#1{\texttt{#1}}\fi
\expandafter\ifx\csname urlprefix\endcsname\relax\def\urlprefix{URL }\fi
\providecommand{\bibinfo}[2]{#2}
\providecommand{\eprint}[2][]{\url{#2}}

\bibitem[{\citenamefont{Cook}(2000)}]{cook00}
\bibinfo{author}{\bibfnamefont{G.}~\bibnamefont{Cook}},
  \bibinfo{journal}{Living reviews in relativity} \textbf{\bibinfo{volume}{3}},
  \bibinfo{pages}{5} (\bibinfo{year}{2000}), \bibinfo{note}{an online version
  can be found at {\tt http://relativity.livingreviews.org}}.

\bibitem[{\citenamefont{Gourgoulhon et~al.}(2002)\citenamefont{Gourgoulhon,
  Grandcl\'ement, and Bonazzola}}]{ggba}
\bibinfo{author}{\bibfnamefont{E.}~\bibnamefont{Gourgoulhon}},
  \bibinfo{author}{\bibfnamefont{P.}~\bibnamefont{Grandcl\'ement}},
  \bibnamefont{and}
  \bibinfo{author}{\bibfnamefont{S.}~\bibnamefont{Bonazzola}},
  \bibinfo{journal}{Phys. Rev. D.} \textbf{\bibinfo{volume}{65}},
  \bibinfo{pages}{044020} (\bibinfo{year}{2002}).

\bibitem[{\citenamefont{Grandcl\'ement
  et~al.}(2002)\citenamefont{Grandcl\'ement, Gourgoulhon, and
  Bonazzola}}]{ggbb}
\bibinfo{author}{\bibfnamefont{P.}~\bibnamefont{Grandcl\'ement}},
  \bibinfo{author}{\bibfnamefont{E.}~\bibnamefont{Gourgoulhon}},
  \bibnamefont{and}
  \bibinfo{author}{\bibfnamefont{S.}~\bibnamefont{Bonazzola}},
  \bibinfo{journal}{Phys. Rev. D.} \textbf{\bibinfo{volume}{65}},
  \bibinfo{pages}{044021} (\bibinfo{year}{2002}).

\bibitem[{\citenamefont{Tichy et~al.}(2003)\citenamefont{Tichy, Br\"{u}gmann,
  and Laguna}}]{tichy03}
\bibinfo{author}{\bibfnamefont{W.}~\bibnamefont{Tichy}},
  \bibinfo{author}{\bibfnamefont{B.}~\bibnamefont{Br\"{u}gmann}},
  \bibnamefont{and} \bibinfo{author}{\bibfnamefont{P.}~\bibnamefont{Laguna}},
  \bibinfo{journal}{Phys. Rev. D} \textbf{\bibinfo{volume}{68}},
  \bibinfo{pages}{064008} (\bibinfo{year}{2003}).

\bibitem[{\citenamefont{Tichy and Br\"{u}gmann}(2004)}]{tichy04}
\bibinfo{author}{\bibfnamefont{W.}~\bibnamefont{Tichy}} \bibnamefont{and}
  \bibinfo{author}{\bibfnamefont{B.}~\bibnamefont{Br\"{u}gmann}},
  \bibinfo{journal}{Phys.Rev. D} \textbf{\bibinfo{volume}{69}},
  \bibinfo{pages}{024006} (\bibinfo{year}{2004}).

\bibitem[{\citenamefont{Yo et~al.}(2004)\citenamefont{Yo, Cook, Shapiro, and
  Baumgarte}}]{yo04}
\bibinfo{author}{\bibfnamefont{H.-J.} \bibnamefont{Yo}},
  \bibinfo{author}{\bibfnamefont{J.}~\bibnamefont{Cook}},
  \bibinfo{author}{\bibfnamefont{S.}~\bibnamefont{Shapiro}}, \bibnamefont{and}
  \bibinfo{author}{\bibfnamefont{T.}~\bibnamefont{Baumgarte}},
  \bibinfo{journal}{Phys. Rev. D} \textbf{\bibinfo{volume}{70}},
  \bibinfo{pages}{084033} (\bibinfo{year}{2004}),
  \bibinfo{note}{gr-qc/0406020}.

\bibitem[{\citenamefont{Cook and Pfeiffer}(2004)}]{cook04}
\bibinfo{author}{\bibfnamefont{G.}~\bibnamefont{Cook}} \bibnamefont{and}
  \bibinfo{author}{\bibfnamefont{H.}~\bibnamefont{Pfeiffer}},
  \bibinfo{journal}{Phys. Rev. D} \textbf{\bibinfo{volume}{70}},
  \bibinfo{pages}{104016} (\bibinfo{year}{2004}),
  \bibinfo{note}{gr-qc/0407078}.

\bibitem[{\citenamefont{{York, Jr.}}(1979)}]{york79}
\bibinfo{author}{\bibfnamefont{J.}~\bibnamefont{{York, Jr.}}}, in
  \emph{\bibinfo{booktitle}{Sources of gravitational radiation}}, edited by
  \bibinfo{editor}{\bibfnamefont{L.}~\bibnamefont{Smarr}}
  (\bibinfo{publisher}{Cambridge University Press},
  \bibinfo{address}{Cambridge}, \bibinfo{year}{1979}), pp.
  \bibinfo{pages}{83--126}.

\bibitem[{\citenamefont{Bowen and {York, Jr.}}(1980)}]{bowen80}
\bibinfo{author}{\bibfnamefont{J.}~\bibnamefont{Bowen}} \bibnamefont{and}
  \bibinfo{author}{\bibfnamefont{J.}~\bibnamefont{{York, Jr.}}},
  \bibinfo{journal}{Phys. Rev. D} \textbf{\bibinfo{volume}{21}},
  \bibinfo{pages}{2047} (\bibinfo{year}{1980}).

\bibitem[{\citenamefont{Cook}(1994)}]{cook94}
\bibinfo{author}{\bibfnamefont{G.}~\bibnamefont{Cook}}, \bibinfo{journal}{Phys.
  Rev. D} \textbf{\bibinfo{volume}{50}}, \bibinfo{pages}{5025}
  (\bibinfo{year}{1994}).

\bibitem[{\citenamefont{Brandt and Br\"ugmann}(1997)}]{brandt97}
\bibinfo{author}{\bibfnamefont{S.}~\bibnamefont{Brandt}} \bibnamefont{and}
  \bibinfo{author}{\bibfnamefont{B.}~\bibnamefont{Br\"ugmann}},
  \bibinfo{journal}{Phys. Rev. Lett.} \textbf{\bibinfo{volume}{78}},
  \bibinfo{pages}{3606} (\bibinfo{year}{1997}).

\bibitem[{\citenamefont{Baumgarte}(2000)}]{baum00}
\bibinfo{author}{\bibfnamefont{T.}~\bibnamefont{Baumgarte}},
  \bibinfo{journal}{Phys. Rev. D.} \textbf{\bibinfo{volume}{62}},
  \bibinfo{pages}{024018} (\bibinfo{year}{2000}).

\bibitem[{\citenamefont{{York, Jr.}}(1999)}]{york98}
\bibinfo{author}{\bibfnamefont{J.}~\bibnamefont{{York, Jr.}}},
  \bibinfo{journal}{Phys. Rev. Lett.} \textbf{\bibinfo{volume}{82}},
  \bibinfo{pages}{1350} (\bibinfo{year}{1999}).

\bibitem[{\citenamefont{Pfeiffer et~al.}(2002)\citenamefont{Pfeiffer, Cook, and
  Teukolsky}}]{pfeiffer02}
\bibinfo{author}{\bibfnamefont{H.}~\bibnamefont{Pfeiffer}},
  \bibinfo{author}{\bibfnamefont{G.}~\bibnamefont{Cook}}, \bibnamefont{and}
  \bibinfo{author}{\bibfnamefont{S.}~\bibnamefont{Teukolsky}},
  \bibinfo{journal}{Phys. Rev. D.} \textbf{\bibinfo{volume}{66}},
  \bibinfo{pages}{024047} (\bibinfo{year}{2002}).

\bibitem[{\citenamefont{Cook}(2002)}]{cook01}
\bibinfo{author}{\bibfnamefont{G.}~\bibnamefont{Cook}}, \bibinfo{journal}{Phys.
  Rev. D} \textbf{\bibinfo{volume}{65}}, \bibinfo{pages}{084003}
  (\bibinfo{year}{2002}).

\bibitem[{\citenamefont{Baumgarte and Shapiro}(2003)}]{baum03}
\bibinfo{author}{\bibfnamefont{T.}~\bibnamefont{Baumgarte}} \bibnamefont{and}
  \bibinfo{author}{\bibfnamefont{S.}~\bibnamefont{Shapiro}},
  \bibinfo{journal}{Phys. Rep.} \textbf{\bibinfo{volume}{376}},
  \bibinfo{pages}{41} (\bibinfo{year}{2003}).

\bibitem[{\citenamefont{Hannam et~al.}(2003)\citenamefont{Hannam, Cook,
  Baumgarte, and Evans}}]{hannam03}
\bibinfo{author}{\bibfnamefont{M.~D.} \bibnamefont{Hannam}},
  \bibinfo{author}{\bibfnamefont{G.~B.} \bibnamefont{Cook}},
  \bibinfo{author}{\bibfnamefont{T.~W.} \bibnamefont{Baumgarte}},
  \bibnamefont{and} \bibinfo{author}{\bibfnamefont{C.~R.} \bibnamefont{Evans}},
  \bibinfo{journal}{Phys. Rev. D} \textbf{\bibinfo{volume}{68}},
  \bibinfo{pages}{064003} (\bibinfo{year}{2003}).

\bibitem[{\citenamefont{Alcubierre et~al.}(2003)\citenamefont{Alcubierre,
  Bruegmann, Diener, Koppitz, Pollney, Seidel, and Takahashi}}]{alcubierre03}
\bibinfo{author}{\bibfnamefont{M.}~\bibnamefont{Alcubierre}},
  \bibinfo{author}{\bibfnamefont{B.}~\bibnamefont{Bruegmann}},
  \bibinfo{author}{\bibfnamefont{P.}~\bibnamefont{Diener}},
  \bibinfo{author}{\bibfnamefont{M.}~\bibnamefont{Koppitz}},
  \bibinfo{author}{\bibfnamefont{D.}~\bibnamefont{Pollney}},
  \bibinfo{author}{\bibfnamefont{E.}~\bibnamefont{Seidel}}, \bibnamefont{and}
  \bibinfo{author}{\bibfnamefont{R.}~\bibnamefont{Takahashi}},
  \bibinfo{journal}{Phys.Rev. D} \textbf{\bibinfo{volume}{67}},
  \bibinfo{pages}{084023} (\bibinfo{year}{2003}).

\bibitem[{\citenamefont{Br\"ugmann et~al.}(2004)\citenamefont{Br\"ugmann,
  Tichy, and Jansen}}]{brugmann04}
\bibinfo{author}{\bibfnamefont{B.}~\bibnamefont{Br\"ugmann}},
  \bibinfo{author}{\bibfnamefont{W.}~\bibnamefont{Tichy}}, \bibnamefont{and}
  \bibinfo{author}{\bibfnamefont{N.}~\bibnamefont{Jansen}},
  \bibinfo{journal}{Phys.Rev.Lett.} \textbf{\bibinfo{volume}{92}},
  \bibinfo{pages}{211101} (\bibinfo{year}{2004}).

\bibitem[{\citenamefont{Imbiriba et~al.}(2004)\citenamefont{Imbiriba, Baker,
  Choi, Centrella, Fiske, Brown, van Meter, and Olson}}]{imbiriba04}
\bibinfo{author}{\bibfnamefont{B.}~\bibnamefont{Imbiriba}},
  \bibinfo{author}{\bibfnamefont{J.}~\bibnamefont{Baker}},
  \bibinfo{author}{\bibfnamefont{D.}~\bibnamefont{Choi}},
  \bibinfo{author}{\bibfnamefont{J.}~\bibnamefont{Centrella}},
  \bibinfo{author}{\bibfnamefont{D.}~\bibnamefont{Fiske}},
  \bibinfo{author}{\bibfnamefont{D.}~\bibnamefont{Brown}},
  \bibinfo{author}{\bibfnamefont{J.}~\bibnamefont{van Meter}},
  \bibnamefont{and} \bibinfo{author}{\bibfnamefont{K.}~\bibnamefont{Olson}},
  \bibinfo{journal}{Phys. Rev. D} \textbf{\bibinfo{volume}{70}},
  \bibinfo{pages}{124025} (\bibinfo{year}{2004}),
  \bibinfo{note}{gr-qc/0403048}.

\bibitem[{\citenamefont{Laguna}(2004)}]{laguna04}
\bibinfo{author}{\bibfnamefont{P.}~\bibnamefont{Laguna}},
  \bibinfo{journal}{Phys. Rev. D} \textbf{\bibinfo{volume}{69}},
  \bibinfo{pages}{104020} (\bibinfo{year}{2004}).

\bibitem[{\citenamefont{Arnowitt et~al.}(1962)\citenamefont{Arnowitt, Deser,
  and Misner}}]{adm}
\bibinfo{author}{\bibfnamefont{R.}~\bibnamefont{Arnowitt}},
  \bibinfo{author}{\bibfnamefont{S.}~\bibnamefont{Deser}}, \bibnamefont{and}
  \bibinfo{author}{\bibfnamefont{C.}~\bibnamefont{Misner}}, in
  \emph{\bibinfo{booktitle}{Gravitation: An Introduction to Current Research}},
  edited by \bibinfo{editor}{\bibfnamefont{L.}~\bibnamefont{Witten}}
  (\bibinfo{publisher}{Wiley}, \bibinfo{address}{New York},
  \bibinfo{year}{1962}).

\bibitem[{\citenamefont{Lichnerowicz}(1944)}]{lich44}
\bibinfo{author}{\bibfnamefont{A.}~\bibnamefont{Lichnerowicz}},
  \bibinfo{journal}{J. Math. Pures at Appl.} \textbf{\bibinfo{volume}{23}},
  \bibinfo{pages}{37} (\bibinfo{year}{1944}).

\bibitem[{\citenamefont{{York, Jr.}}(1971)}]{york71}
\bibinfo{author}{\bibfnamefont{J.}~\bibnamefont{{York, Jr.}}},
  \bibinfo{journal}{Phys. Rev. Lett.} \textbf{\bibinfo{volume}{26}},
  \bibinfo{pages}{1656} (\bibinfo{year}{1971}).

\bibitem[{\citenamefont{{York, Jr.}}(1972)}]{york72}
\bibinfo{author}{\bibfnamefont{J.}~\bibnamefont{{York, Jr.}}},
  \bibinfo{journal}{Phys. Rev. Lett.} \textbf{\bibinfo{volume}{28}},
  \bibinfo{pages}{1082} (\bibinfo{year}{1972}).

\bibitem[{\citenamefont{Pfeiffer and {York, Jr.}}(2003)}]{york03}
\bibinfo{author}{\bibfnamefont{H.}~\bibnamefont{Pfeiffer}} \bibnamefont{and}
  \bibinfo{author}{\bibfnamefont{J.}~\bibnamefont{{York, Jr.}}},
  \bibinfo{journal}{Phys. Rev. D} \textbf{\bibinfo{volume}{67}},
  \bibinfo{pages}{044022} (\bibinfo{year}{2003}).

\bibitem[{\citenamefont{Hannam}(2003)}]{mdhPhD}
\bibinfo{author}{\bibfnamefont{M.~D.} \bibnamefont{Hannam}}, Ph.D. thesis,
  \bibinfo{school}{University of North Carolina at Chapel Hill}
  (\bibinfo{year}{2003}).

\bibitem[{\citenamefont{Brill and Lindquist}(1963)}]{brill63}
\bibinfo{author}{\bibfnamefont{D.}~\bibnamefont{Brill}} \bibnamefont{and}
  \bibinfo{author}{\bibfnamefont{R.}~\bibnamefont{Lindquist}},
  \bibinfo{journal}{Phys. Rev.} \textbf{\bibinfo{volume}{131}},
  \bibinfo{pages}{471} (\bibinfo{year}{1963}).

\bibitem[{\citenamefont{Misner and Wheeler}(1957)}]{misner57}
\bibinfo{author}{\bibfnamefont{C.}~\bibnamefont{Misner}} \bibnamefont{and}
  \bibinfo{author}{\bibfnamefont{J.}~\bibnamefont{Wheeler}},
  \bibinfo{journal}{Ann. Phys.} \textbf{\bibinfo{volume}{2}},
  \bibinfo{pages}{525} (\bibinfo{year}{1957}).

\bibitem[{\citenamefont{O'Murchadha}(1992)}]{omurch92}
\bibinfo{author}{\bibfnamefont{N.}~\bibnamefont{O'Murchadha}}, in
  \emph{\bibinfo{booktitle}{Approaches to Numerical Relativity}}, edited by
  \bibinfo{editor}{\bibfnamefont{R.}~\bibnamefont{D'Inverno}}
  (\bibinfo{publisher}{Cambridge University Press},
  \bibinfo{address}{Cambridge}, \bibinfo{year}{1992}), pp.
  \bibinfo{pages}{83--93}.

\bibitem[{cac()}]{cactus}
\bibinfo{note}{{\tt http://www.cactuscode.org}}.

\bibitem[{\citenamefont{O'Murchadha and {York, Jr.}}(1974)}]{omurch74c}
\bibinfo{author}{\bibfnamefont{N.}~\bibnamefont{O'Murchadha}} \bibnamefont{and}
  \bibinfo{author}{\bibfnamefont{J.}~\bibnamefont{{York, Jr.}}},
  \bibinfo{journal}{Phys. Rev. D.} \textbf{\bibinfo{volume}{10}},
  \bibinfo{pages}{2345} (\bibinfo{year}{1974}).

\bibitem[{\citenamefont{Dettwyler}(2004)}]{mhdMS}
\bibinfo{author}{\bibfnamefont{M.}~\bibnamefont{Dettwyler}}, Master's thesis,
  \bibinfo{school}{University of Texas at Brownsville} (\bibinfo{year}{2004}).

\end{thebibliography}

\end{document}